# Beyond Profiling: Scaling Profiling Data Usage to Multiple Applications

Chris Quackenbush
Google
cquackenbush@gmail.com

Mohamed Zahran
CS Department, NYU
mzahran@cs.nyu.edu

*Abstract*—Profiling techniques are used extensively at different parts of the computing stack to achieve many goals. One major goal is to make a piece of software execute more efficiently on a specific hardware platform, where efficiency spans criteria such as power, performance, resource requirements, etc. Researchers, both in academia and industry, have introduced many techniques to gather, and make use of, profiling data. However, one thing remains unchanged: making application A run more efficiently on machine 1. In this paper, we extend this criteria by asking: can profiling information of application A on machine 1 be used to make application B run more efficiently on machine 1? If so, then this means as machine 1 continues to execute more applications, it becomes better and more efficient.

We present a generalized method for using profiling information gathered from the execution of programs from a limited corpus of applications to improve the performance of software from outside our corpus. As a proof of concept, we apply our technique to the specific problem of selecting the most efficient last-level-cache with which to execute an application. We were able to turn off an average of 19% of last-level-cache blocks for selected programs from PARSEC benchmark suite and only saw an average 2.8% increase in the rate of last-level cache misses.

## I. INTRODUCTION

There are many techniques for gathering and utilizing profiling data, but they all share a common constraint: each software application must be instrumented and executed on a hardware platform at least once to gather profiling information before any useful optimizations can be made. We propose that there is a finite set of patterns along which hardware/software interactions can occur to give best performance. For example, given a cache configuration, there are finite set of memory access patterns that yield low cache misses. Or, given a memory access pattern, we can build the best cache configuration that yields the lowest number of misses. We can determine the best hardware configuration for each interaction pattern by capturing profiling data from a corpus of representative applications on a small set of hardware configurations. We can then build a system which recognizes these patterns while executing applications from outside of the corpus and uses the most performant hardware configuration. The more profiling data is available, the more patterns will be recognized and the better the execution.

This method can be compared with other techniques for optimizing performance during application execution. Optimizing compilers are one of the most common forms of performance enhancement, but these techniques are forced to be very general because the actual program behavior at runtime is unknown [10]. Profile-guided compilation [19] uses data from previous executions to select more appropriate optimizations, but requires the application to be run and profiled at least once before any performance benefits can be realized [19]. More broadly, all techniques that require the program to be recompiled add additional overhead before running the application.

Other feedback-directed optimizations such as branch prediction [22] can be done at runtime, but suffer from the weakness that these optimizations must be based on the behavior of the application in its earlier phase of execution. In many programs, the same subroutine or basic block may have different behavior over the course of execution [19] depending on the data being processed.

Using the method presented in this paper, profiling data from previous applications can be reused without ever profiling the performance of a target application, but only observing its interaction pattern with the hardware. This allows us to provide profile-guided optimizations to all applications the first time they are executed. Moreover, as we collect more profiling information, our presented method can become more and more precise.

## II. RELATED WORK

Previous work on improving performance through profiling has been divided into two distinct categories: those techniques that modify the hardware platform and those that modify the application software or software platform. Chen et al [14] used supervised learning to predict detailed software profiles from coarse granularity hardware event sampling data (e.g. L2 cache write misses). Han and Abdelrahman [6] used a machine learning model trained on benchmarks to successfully predict when local memory usage in GPUs would be an effective optimization. Ipek et al [7] used a similar technique to choose hardware configurations in an offline way. They presented a method for using sampled hardware performance and an artificial neural network to predict performance of unseen hardware designs so as to efficiently explore of the architectural design space when creating multiprocessors.

Other methods for reconfiguring hardware online as a program executes which make use of feedback-directed optimization techniques, but do not use machine learning have also appeared. Abella et al [8] presented a heuristic for predicting the last instruction accessing each L2 cache block and disabling the blocks to reduce power consumption. Banerjee, Subhasis, and Nandy [9] turned off entire cache ways based on the observed miss rate as an application executes.

Using profiling information from previous runs of the *same* program to produce better future executions without recompiling the application is not a new technique. Schulte et al [13] presented a method for using genetic algorithms to continually profile and transform application object code to increase power efficiency across multiple executions. Work has also been done to use profiling information from applications to reconfigure the execution environment. Yuan, Guo, and Chen [12] used profile-guided optimization to recompile the Linux kernel (as opposed to the application software) to produce a more performant environment for user software.

## III. PROPOSED IDEA

**We present a method for using profiling information gathered from the execution of programs from a limited corpus of applications to improve the performance of software from outside our corpus**. That is, we answer this question: how can profiling information from a small set of known programs be used to improve the performance of all unknown programs? As shown in Figure 1, our technique is performed in two phases: a training phase that occurs once and produces a model of all profiling data from the training corpus and an execution phase that occurs while the target application is running and reconfigures the system to the optimal state for executing the current program.

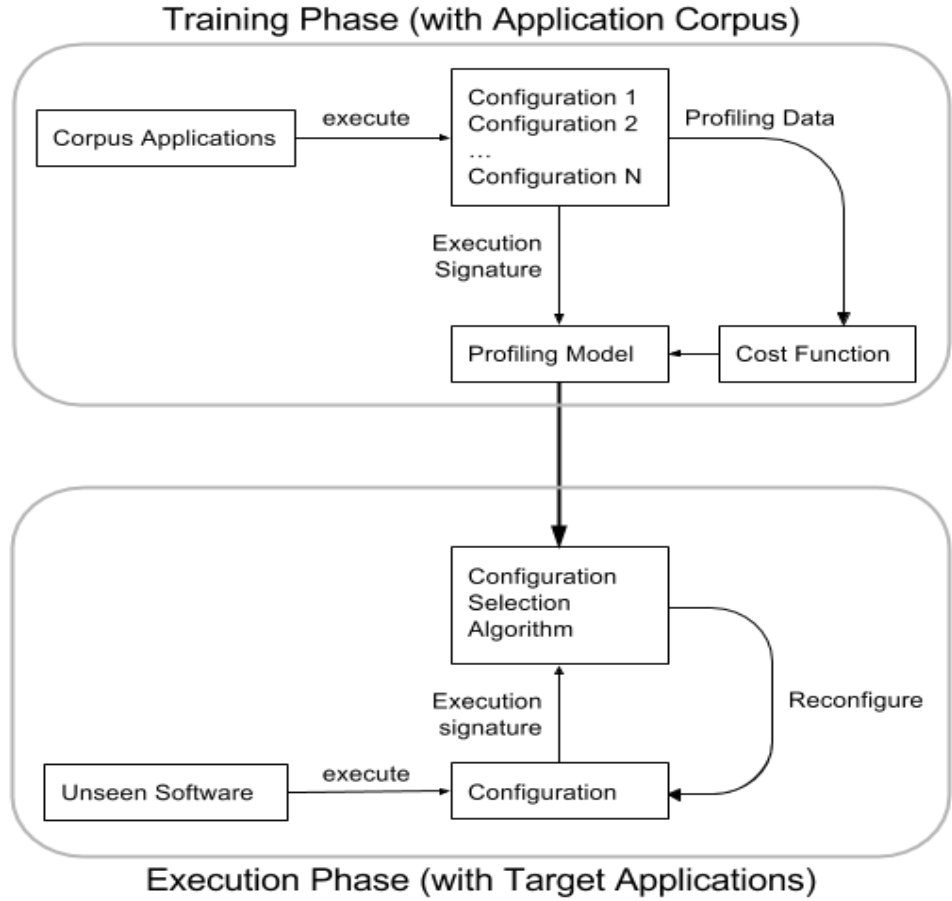

**Figure 1: An overview of the methodology presented in this paper. The training phase occurs only once and the execution phase happens each time an application runs.**

During the training phase, profiling information is collected as each corpus application executes in each system configuration. During each execution, signatures summarizing the hardware/software interaction pattern are also collected. These interactions could include cache behavior, instruction pipeline behavior, memory behavior, etc. These signatures are combined with the profiling data to build a model which maps each observed execution signature to an optimal system configuration (as determined by the profiling data). The job of the model is to predict the optimal configuration from an observed signature. A model could be a simple as a hash map of signatures or as complicated as a neural network.

During the execution phase, signatures are collected continually as the hardware/software interaction patterns occur. These signatures are consumed by a configuration selection algorithm which uses the model of profiling data created in the training phase to predict which system configuration will be most performant. When a new configuration is predicted to be more performant than the current configuration, the system is immediately reconfigured to the state selected by the algorithm. This process continues until the target application ends.

## IV. RECONFIGURING THE HARDWARE PLATFORM

To illustrate our method, we provide a system to perform the following task: Given a running candidate application and predefined set of hardware configurations, select the best hardware configuration with which to execute the application. This is repeated at each phase [1] during application execution. We first define a cost function which will specify the hardware configuration that performs best. By selecting the configuration that minimizes this function we will make the hardware platform more performant along the dimension of the cost function. The cost function can measure execution time, power consumption, resource utilization, or any other chosen metric.

During the training phase, each application from the corpus will be executed on each predefined hardware configuration[1]. The resulting profiling data will be analyzed with respect to the cost function to determine which hardware configuration is most efficient and a model will be built mapping the observed signatures to that hardware state. As the target application executes, that model is used to predict the optimal hardware configuration and the hardware platform is continually reconfigured to the state determined to be most efficient for the current application behavior (as summarized by the most recent execution signature).

A. *Definitions*

1) *Program Phases and Working Sets*

Application software typically exhibits long sequences of uniform behavior interrupted by abrupt periods of unstable activity [15]. These periods of stability are called program phases [23]. We can divide each phase into periods of equal length (as measured by any regularly occurring event, such as CPU cycles, instructions committed, memory accesses, etc). We will call each of these periods an execution window. The working set for each execution window is defined as the set of distinct resources accessed during that window [20]. The segments are typically memory regions of some fixed size such as a cache row or a basic block.

Program phases are primarily caused by working sets and the abrupt phase change is a reflection of a change in the working set [1]. The working set of instructions is defined by the control passing through subroutines and nested loops. The set of instructions executed is very similar in some intervals and suddenly different as control exits from a loop or subroutine [11].

2) *Working Set Signatures*

A working set signature (WSS) is a compressed representation of the entire working set over a single window. An instructions WSS is related to cache performance measurements (e.g. miss rate), but because it is independent of the hardware interaction it doesn't capture details of the specific memory implementation or geometry [5]. A memory WSS relates directly to the blocks or pages accessed during the window and can therefore capture more details of the software/hardware interaction.

$$\frac{hamming\_weight(WSS1 \oplus WSS2)}{hamming\_weight(WSS1 \mid WSS2)} \quad (1)$$

The distance function between two WSS is shown in Equation (1). The WSS distance function counts the number of one bits in the bitwise exclusive or of two signatures and the one bits of the bitwise inclusive or and expresses the ratio of the two counts. The *hamming_weight* function is the sum of all the bits in its argument (i.e. the number of one bits).

3) *Memory Access Signatures*

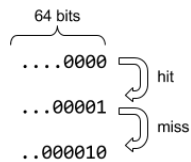

**Figure 2: The memory access signature**

A memory access signature (MAS) is a record of the recent performance for a unit of cache memory. As illustrated in Figure 2, the MAS is 64 bits wide and shifted left each time the cache unit is accessed. When the access results in a cache hit, the least significant bit is set to one; on a cache miss, the least significant bit is set to zero. A MAS can be computed for any unit of cache memory (e.g. a block, a set, a way, etc).

B. *Experimental Considerations*

As will be described in Section 5, our implementation will be reconfiguring last-level cache and we will make extensive use of a MAS for the entire last-level cache. When designing a proof-of-concept system to reconfigure the hardware platform we had to consider if the MAS was an appropriate signature for our implementation. Figures 3 and 4 show the memory access pattern across for the same application across multiple LLC sizes.

---

[1] The method presented in this paper can be used with any hardware aspect (we used LLC size as an example), such as: branch prediction table size, branch prediction algorithm, scheduling of instructions for execution, cache replacement policy, etc. The design space is manageable for each of these aspects, and since it is done offline then there is no performance loss.

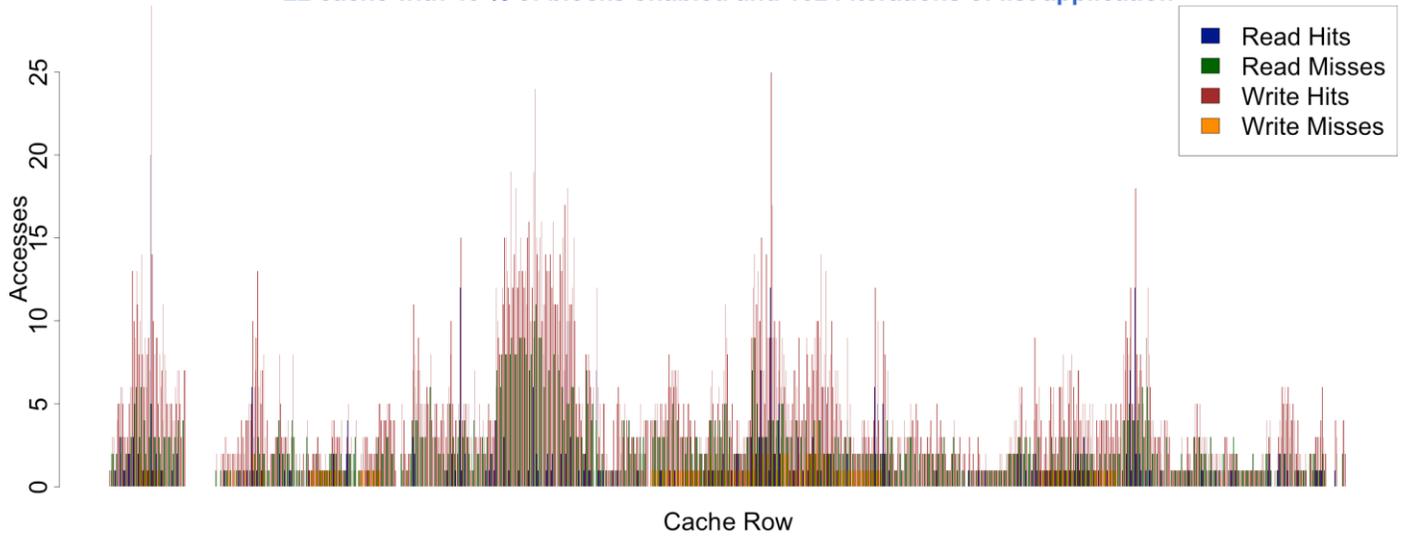

Figure 3: The memory access pattern (read/write hits/misses) for each cache row when running the "list" application from our corpus with 40% of the LLC blocks turned on.

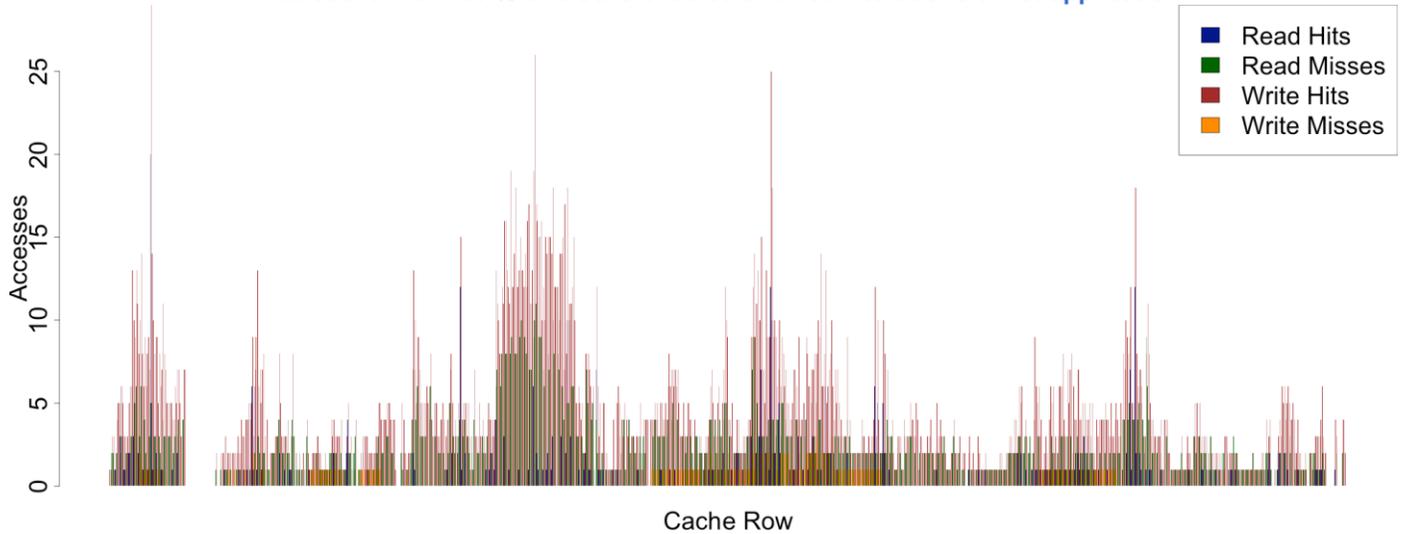

Figure 4: The memory access pattern with 100% of the LLC turned on.

Figures 3 and 4 show hardly any difference in the cache access patterns. Turning off 60% of the blocks seems to have lead to a barely noticeable increase in read misses and no detectable increase in write misses. This indicates that memory access patterns for an application remain stable even if the size of the LLC changes.

## V. OUR PROOF-OF-CONCEPT IMPLEMENTATION: RECONFIGURING THE LAST-LEVEL CACHE

### A. *The Simulated Hardware Platform*

In the following sections, we use the motivating example of selecting a last-level-cache size from a small set of available cache sizes to minimize the amount of cache turned on during execution hence saving power. This approach can then be analogized to tuning other hardware parameters.

We use the Multi2Sim system simulation framework [17] to simulate the execution of our corpus and candidate applications. The machine we simulated uses a single X86 CPU with a single thread and a memory geometry summarized in Table 1.

**Table 1: The cache configuration of our simulated machine**

| Cache | Ways | Sets | Block Size | Latency |
|---|---|---|---|---|
| Level 1 | 2-way associative | 256 sets | 64 bytes | 2 cycles |
| Level 2 | 16-way associative | 1024 sets | 64 bytes | 10 cycles |

The main memory was configured with 256 byte blocks and a 600 cycle latency.

B. *Reconfigurable Last Level Cache*

We have extended Multi2Sim to support the ability to turn off individual cache blocks in the last level cache (in the case of these experiments the LLC is always the L2 cache). Our system can reduce the number of accessible cache blocks in the LLC to any number greater than the number of sets. To ensure that the entire main memory remains cacheable, this system will never disable all blocks in a set. This constraint is the reason that our system does not allow for the number of blocks in the LLC to be reduced below the number of sets. Otherwise, we will need to change the number of sets, which will make the cache controller more complicated, and hence slower and power hungry.

Algorithm 1 describes the procedure for choosing blocks to turn off. This procedure minimizes the cost of decreasing the cache size by first preferring to disable blocks that have never been accessed and then disabling blocks that have never been written; only once all other blocks have been disabled will it begin to incur the overhead of turning off dirty cache blocks.

*ALGORITHM 1: LAST LEVEL CACHE RECONFIGURATION*

```
FUNCTION (blocks_to_turn_off)
  pass = 0
  WHILE (blocks_to_turn_off > 0)
    pass += 1
    FOREACH way in LLC_ASSOCIATIVITY
      FOREACH set in LLC_ROWS
        block = set[way]
        IF exactly one way of set is turned on
          continue
        IF block is accessed AND pass < 2
          continue
        IF block is dirty AND pass < 3
          continue
        turn off block
        Blocks_to_turn_off -= 1
```

To implement Algorithm 1 in hardware we a table of accessed bits and dirty bits: one for each block in the LLC. Each cache row also maintains a count of the number of blocks that are turned on in that row and no row is ever allowed to turn off all its blocks. The cache blocks are turned off in three passes and in each pass all blocks are turned off simultaneously. The first pass turns off blocks that are marked in neither table (up to the total number of blocks to turn off). If a second pass is needed, blocks that are not marked in the dirty table are turned off. The third pass turns off any remaining blocks. The LLC will never disable more than the required number of blocks and blocks with a lower way index are preferred to spread the disabled blocks across all the LLC sets.

Although this algorithm required three rounds, only the third round needs to block pipeline execution (as the dirty blocks are written back to memory). The first two rounds happen in parallel with application execution. In practice, we found that the third round is very rarely needed; we never observed it in any of our experiments.

### C. *Cache Size Selection Algorithms*

We have implemented three algorithms in Multi2Sim for selecting the the number of blocks to turn off in the last-level cache: the extended working set signatures algorithm [2], one that uses a bloom filter [21] as the profiling model (BLOOM), and one that uses an artificial neural network [18] as the profiling model (ANN). BLOOM and ANN are new procedures which we implemented to utilize profiling data collected from our application corpus while EWSS makes no use of profiling data. We will use EWSS as a baseline procedure against which to compare our new methodology in addition to the base configuration where each block in the LLC is always turned on.

#### 1) *The Baseline Algorithm*

As a baseline against which to compare our new methodology, we implemented the extended working set signatures method (EWSS) introduced by Dhodapkar and Smith [2] for selecting a cache size. This algorithm iterates over all available hardware configurations and uses online hardware event sampling in each configuration to determine the optimal hardware state. The EWSS algorithm uses the instruction cache working set and is based on earlier work by Balasubramonian et al [3] which was presented for adjusting the L1 cache size. A very similar algorithm was designed by Zhang, Vahid, and Lysecky [4] for tuning cache memory along multiple parameters in embedded systems via an exhaustive search of the hardware configuration space during execution.

EWSS does not make use of any corpus of profiling data and does not have a training phase to build a model for predicting the optimal configuration. Instead, each time a new program phase is entered, EWSS iterates over all available configurations and profiles application behavior in each one; it then reconfigures the system to the state observed to be more performant. As the basis for its program phase detection, EWSS uses a working set signature derived by hashing the address of each instruction fetched from the instruction cache.

ALGORITHM 2: EXTENDED WORKING SET SIGNATURE (EWSS)

```
prev_wss, wss: 128 bytes initialized to zero

During each window of 100,000 instructions:
index = hash of instruction_pointer into the range [0, 1024]
wss[index] = 1

After each 100,000 instruction window:
distance = WSS_DISTANCE(prev_wss, wss)
IF distance > 0.5
  unstable_windows++
  IF unstable_windows > 10
    stable_windows = 0
    set LLC to maximum size
ELSE
  stable_windows++
  IF stable_windows > 4
    unstable_windows = 0
    IF all LLC configurations were tested in the current phase
      set LLC to best configuration
    ELSE
      set LLC to next configuration
prev_wss = wss
wss = 0
```

The EWSS method uses instruction cache WSS distance (Equation 1) to detect stability across program execution windows and reconfigures the LLC each time it detects a new program phase is entered. Once a new phase is detected via stability in the WSS distance the EWSS algorithm iterates over every available size of the LLC (one per window) and selects the most efficient size for the remainder of the program phase. In our case, efficiency is determined by the size with the fewest cache misses.

2) *The Bloom Filter Algorithm*

The first procedure we implemented that takes advantage of profiling data from our corpus is described in Algorithm 3. This method populates a bloom filter [21] for each hardware configuration with every observed memory access signatures (MAS) that performed best with that configuration.

ALGORITHM 3: BLOOM FILTER MEMORY ACCESS SIGNATURES (BLOOM)

```
bloom_filter_list: array of one filter per LLC configuration

Populate bloom filters during offline training:
FOREACH mas in all MAS observed in training corpus
  best_configuration = best observed configuration for MAS
  add_to_filter(mas, bloom_filter_list[best_configuration])

Each time the LLC is accessed:
mas = updated MAS (see Figure 2)
FOREACH configuration in LLC configurations
  IF check_filter(mas, bloom_filter_list[configuration]
    set LLC to configuration
    return
```

As described in Section 6, the corpus of training applications is used to generate a set of MAS and profiling data is used to associate each MAS with an optimal hardware configuration. The model used in the BLOOM algorithm is a set of bloom filters, one corresponding to each hardware configuration. Each observed MAS is added to the bloom filter corresponding to its associated optimal hardware configuration. Figure 5 shows the training phase of BLOOM where the model (a set of bloom filters) is built. In our experiment, the cost function was selected to minimize the LLC size without increasing the number of execution cycles. The implementation of this function will be described in section 6C.

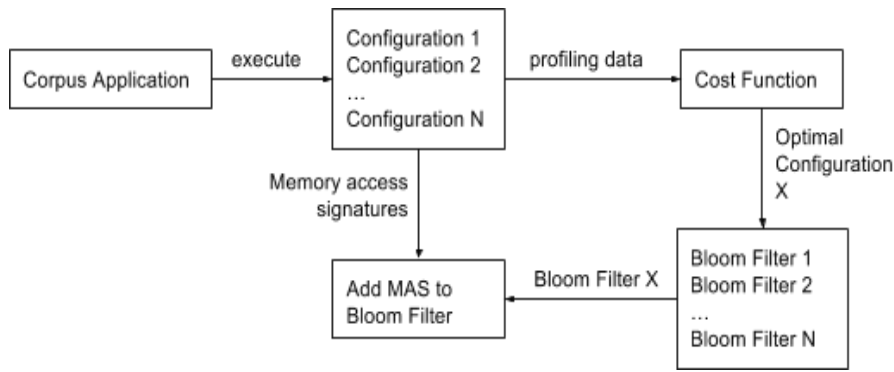

Figure 5: The process for adding MAS to each bloom filter. These steps performed once for each application and data size in the training corpus.

During execution of the candidate software, a MAS is updated each time the LLC is accessed. Membership is tested in every bloom filter for each new MAS and when a MAS belonging to one of the filters is found the LLC is immediately reconfigured to the corresponding size. If the MAS is found in multiple filters the smaller size is preferred. Figure 6 show the execution phase where the model is used to reconfigure the hardware state.

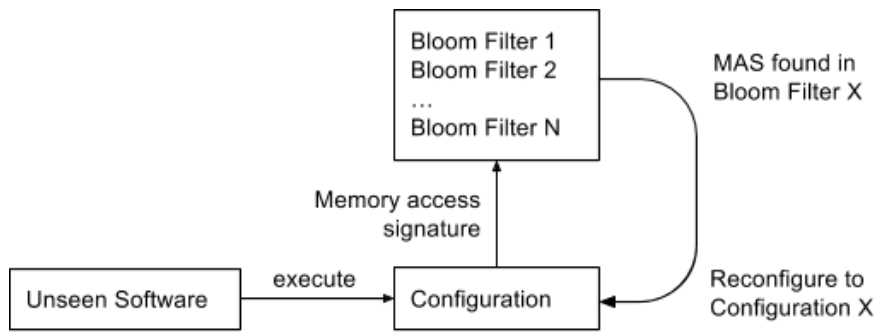

**Figure 6: The process for selecting a LLC size using the bloom filters.**

3) *The Artificial Neural Network Algorithm*

As with the BLOOM algorithm, the Artificial Neural Networks procedure (ANN) described in Algorithm 4 also uses the mapping of MAS to optimal LLC configurations generated by the corpus of training applications. The ANN algorithm trains a neural network to recognize MAS and determine which LLC configuration will be optimal. We used the FANN library [18] to implement artificial neural networks in the training phase and integrated this library with Multi2Sim to control LLC sizing during execution.

The ANN algorithm uses a single neural network with one input neuron for each bit in the MAS and one output neuron for each LLC configuration. These output neurons form the configuration vector which will select which configuration is optimal given an observed MAS. Each bit in this vector corresponds to a distance hardware configuration, so for the output to be valid exactly one bit in the vector must be set and the others must all be zero, otherwise the selected configuration is ambiguous.

Profiling data from the training corpus associates each unique observed MAS with an optimal LLC configuration vector as determined by the cost function described in Section 6C. During the training phase, only perfect matches between input MAS vector and output configuration vectors are considered to be successfully learned. Algorithm 4 uses the hamming_weight function from Equation (1) which is defined as the sum of all the bits in its argument.

ALGORITHM 4: ARTIFICIAL NEURAL NETWORK MEMORY ACCESS SIGNATURES (ANN)

```
fann: a neural network

Train neural network offline with profiling data:
Teach fann MAS → configuration vector
via supervised learning
only perfect matches are considered learned

Each time the LLC is accessed:
mas = updated MAS (see Figure 2)
output_neurons = run(mas, fann)
IF hamming_weight(output_neurons) == 1
  FOREACH configuration in LLC configurations
    IF output_neurons[configuration]
      set LLC to configuration
      return
```

The neural network is taught to learn a function which maps a MAS to a vector of bits with one bit for each hardware configuration. In the training data, all bits of the output vector are zero except for the index corresponding to the optimal LLC configuration. Figure 7 shows the training phase of the ANN algorithm where the artificial neural network model is built.

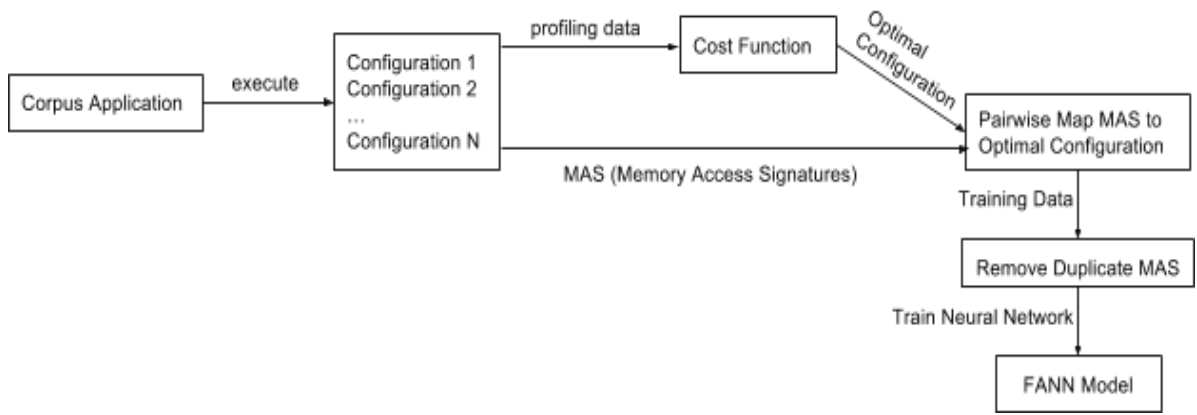

**Figure 7: The process for training a neural network to learn the function mapping MAS to optimal hardware configuration.**

While the candidate software executes, a MAS is updated each time the LLC is accessed. Each new MAS is run through the neural network to obtain an output vector of one bit for each LLC configuration. If none of the bits in the output vector are 1 or if multiple bits are 1 the output is considered to be low confidence and no action is taken. If there is exactly one bit set to 1 in the output vector then the LLC is immediately reconfigured to the size corresponding to that index. Figure 8 shows the execution phase of the ANN algorithm.

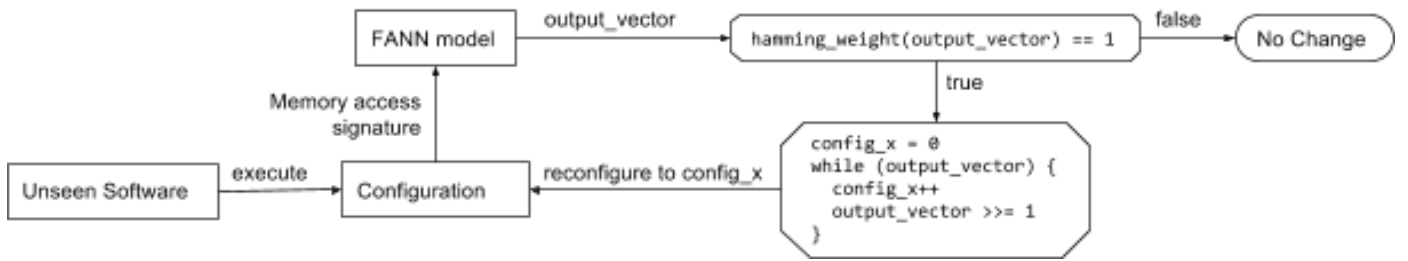

**Figure 8: The process for selecting a LLC size using the neural network.**

# VI. GATHERING PROFILING DATA

## A. *The Application Corpus*

The profiling data used to train the cache size selection algorithms described in Section 5 comes from an application corpus we developed of four small programs listed in Table 2. Each application is focused on a single function and can be executed with input data of three sizes.

**Table 2: The applications of the training corpus used to collect profiling data**

| Application | Description | Data Sizes |
|---|---|---|
| list | Builds a doubly linked list and then traverses it from head to tail | 9, 100, and 1024 nodes |
| sort | Creates an array of randomly generated 32-bit integers and performs quicksort on the data | 9, 100, or 1024 integers |
| transp | Creates an array of randomly generated 32-bit integers and, treating it as a square matrix, populates an identically sized array with the transpose of that matrix. | 9, 100, or 1024 integers interpreted as a 3x3, 10x10, or 32x32 square matrix |
| mul | Creates two arrays of randomly generated 32-bit integers and, treating them as a square matrices, populates an identically sized array with the product of those matrices. | 18, 200, or 2048 integers interpreted as a two 3x3, 10x10, or 32x32 square matrices |

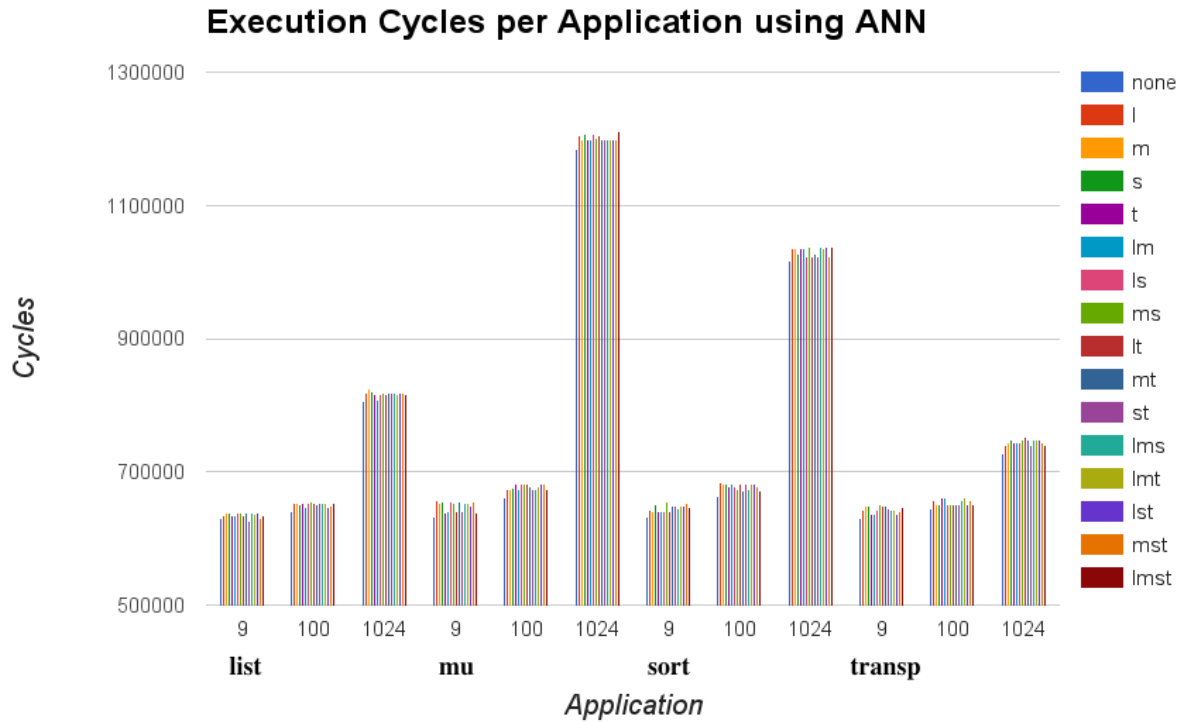

**Figure 9:** Application run time for our training programs when executed with the ANN algorithm using a neural network trained on profiling data from each subset of our application corpus. The legend uses a single letter to indicate each application. The letter "l" indicates the "list" application, "m" is the "mul" application, "s" the "sort" application, and "t" the "transp" application; for example, the bars labeled "lms" trained the neural network on profiling data from the list, mul, and sort applications but not the transp application. The bars labeled "none" use the full LLC size without the ANN algorithm.

Figure 9 shows how application execution time is affected by training the ANN function on different permutations of data from our training corpus. In each case, the execution time is primarily affected by the application and data size. Using profiling data from more or fewer applications does not affect the run time of the application or the performance of the ANN (though it will affect the overall efficiency of each application execution).

B. *Signature Selection*

Each application from the corpus was run at each data size and during each execution the Memory Access Signatures was recorded after each LLC access. This produced 12 sets of MAS. All MAS appearing in multiple sets were considered non-predictive and removed from the training data leaving 12 sets of globally unique MAS each one observed during the execution of a single application on a single data size. Each set of MAS was then associated with a hardware configuration and this mapping from MAS to optimal LLC size was used to populate the bloom filters in Algorithm 3 and the neural network in Algorithm 4.

C. *Associating a Signature with an LLC Size*

The cache size selection algorithms described in Section 5 all require a finite number of hardware configurations to choose among. Theoretically, these configurations could be as fine grained as a single cache block or row. For these experiments, we have chosen to select among 5 different LLC sizes. We chose these sizes to be 20%, 40%, 60%, 80%, and 100% of the maximum LLC size which is 16K blocks of 64 bytes each (1024 KB).

**Table 3: Optimal LLC size selection process for the list application at all three data sizes**

| name | data size | LLC cache percentage | execution cycles | optimal last-level cache size |
|---|---|---|---|---|
| list | 9 | 100% | 627805 | |
| | | 80% | 621713 | |
| | | 60% | 619575 | X |
| | | 40% | 637718 | |
| | | 20% | 685021 | |
| | 100 | 100% | 643294 | |
| | | 80% | 641245 | X |
| | | 60% | 643553 | |
| | | 40% | 655375 | |
| | | 20% | 709587 | |
| | 1024 | 100% | 805544 | |
| | | 80% | 800282 | X |
| | | 60% | 801513 | |
| | | 40% | 807819 | |
| | | 20% | 878314 | |

Each of the 12 application and data size paris from our training corpus was executed with LLC fixed at each of these 5 levels. As shown in Figure 1, our method requires a cost function to evaluate which configuration is most efficient when building a profiling model. In this experiment, we chose a cost function which minimized the LLC size without increasing the execution time. This process of calculating this cost function for the list application is illustrated in Table 3. The set of MAS associated with the optimal application and data size was then mapped in the profiling model to the smallest LLC size that did not increase the execution time as measured in simulated CPU cycles.

## VII. EXPERIMENTAL RESULTS

We used benchmarks from the PARSEC 2.1 [16] benchmark suite as candidate applications to evaluate the efficacy of our LLC size selection method. Figure 12 shows the LLC size selected by the EWSS, BLOOM, and ANN algorithms over the course of executing the blackscholes benchmark from the PARSEC suite.

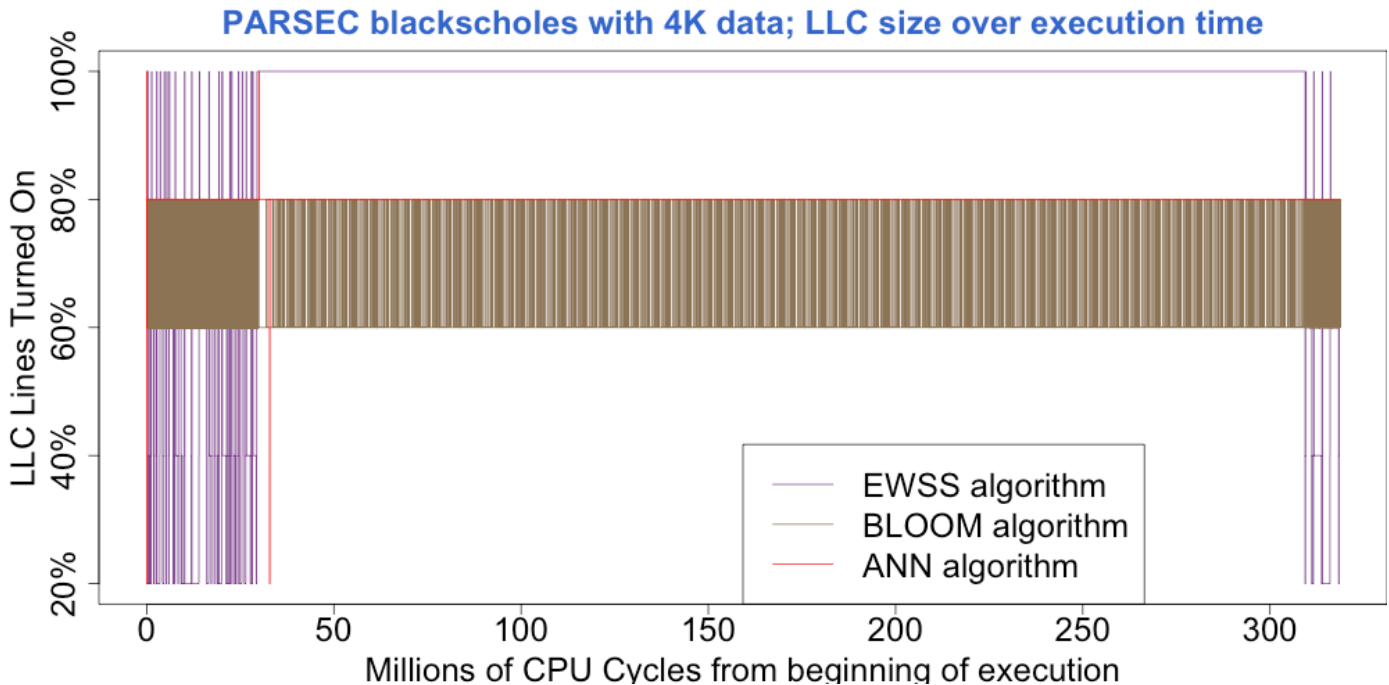

**Figure 12: The function of LLC size over time as determined by the EWSS, BLOOM, and ANN algorithms. The stair-step nature of the EWSS algorithm can be clearly observed near the start of execution as it iterates through all configurations after each program phase change.**

As shown in Figure 12, the ANN and BLOOM algorithms alternate between powering off 20% and 40% of the LLC for the majority of the execution period (although the BLOOM algorithm alternates much more frequently). Due to working set instability, the WSS algorithm barely performs better than operating without any hardware adjustments.

In this example, the ANN algorithm powered an average of 80% of the LLC, the BLOOM algorithm powered 61% of the LLC on average, and and EWSS algorithm powered on average 94% of the LLC. At the same time, the LLC miss rate was 2.8% while using EWSS, 3.1% while using BLOOM, and 3.0% when using ANN. However, there was less that 1% difference in execution time among the three algorithms. This makes ANN and BLOOM an improvement of 15% and 35% in LLC cache size respectively at the cost of only ¼ percentage points additional miss rate.

We performed this analysis on five different benchmarks from the PARSEC suite and the results are summarized in Figures 10 through 12.

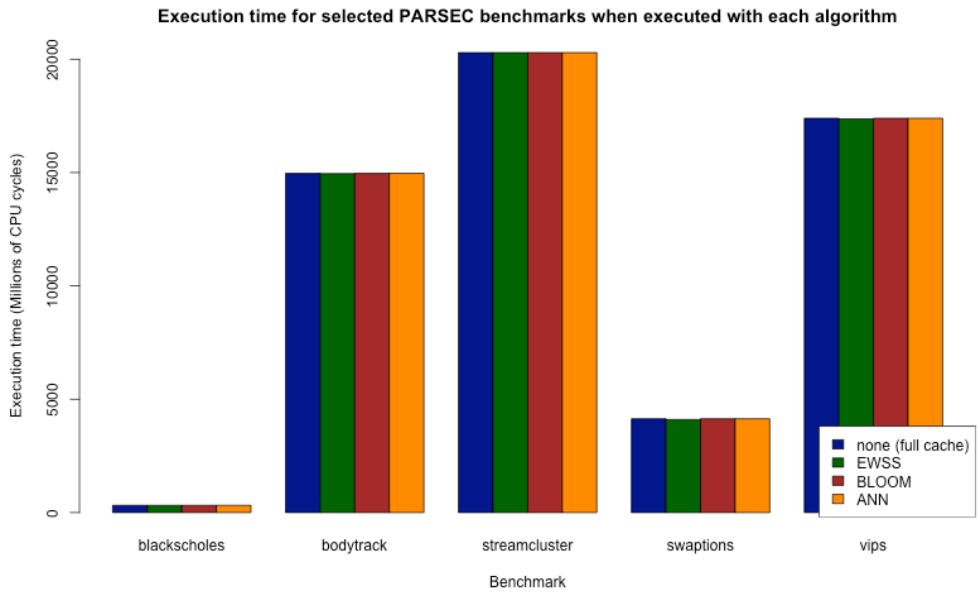

**Figure 10:** Execution time of selected benchmarks from the PARSEC suite using the LLC size selection algorithms presented in this paper. The total number of cycles is not significantly affected by the algorithms.

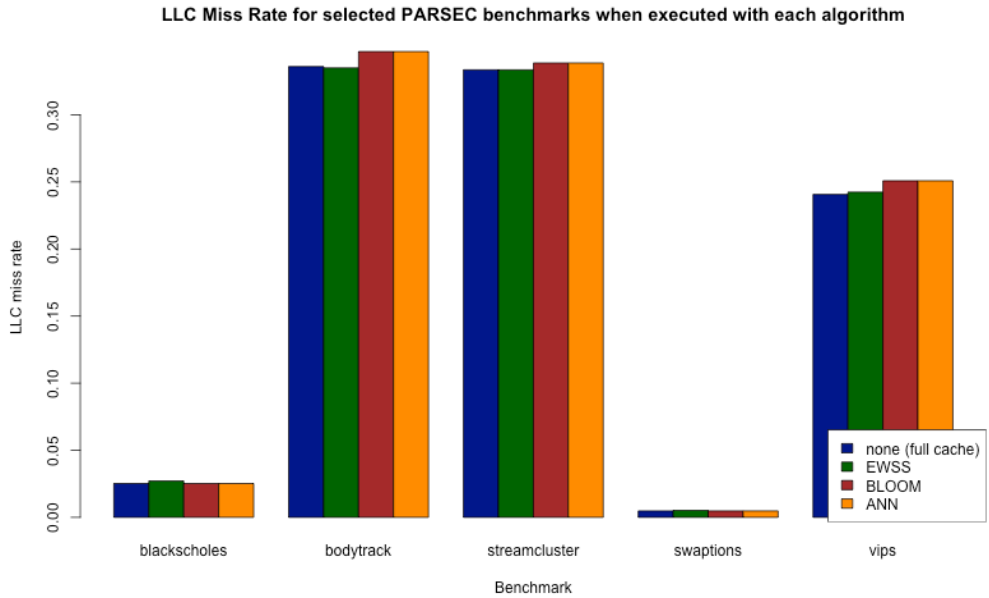

**Figure 11:** LLC miss rate of selected benchmarks from the PARSEC suite using the size selection algorithms presented in this paper. Reducing the cache size does result in a small increase in the frequency of a cache miss. An average increase of 2.8% was observed in both the BLOOM and ANN algorithm when compared to the full cache size.

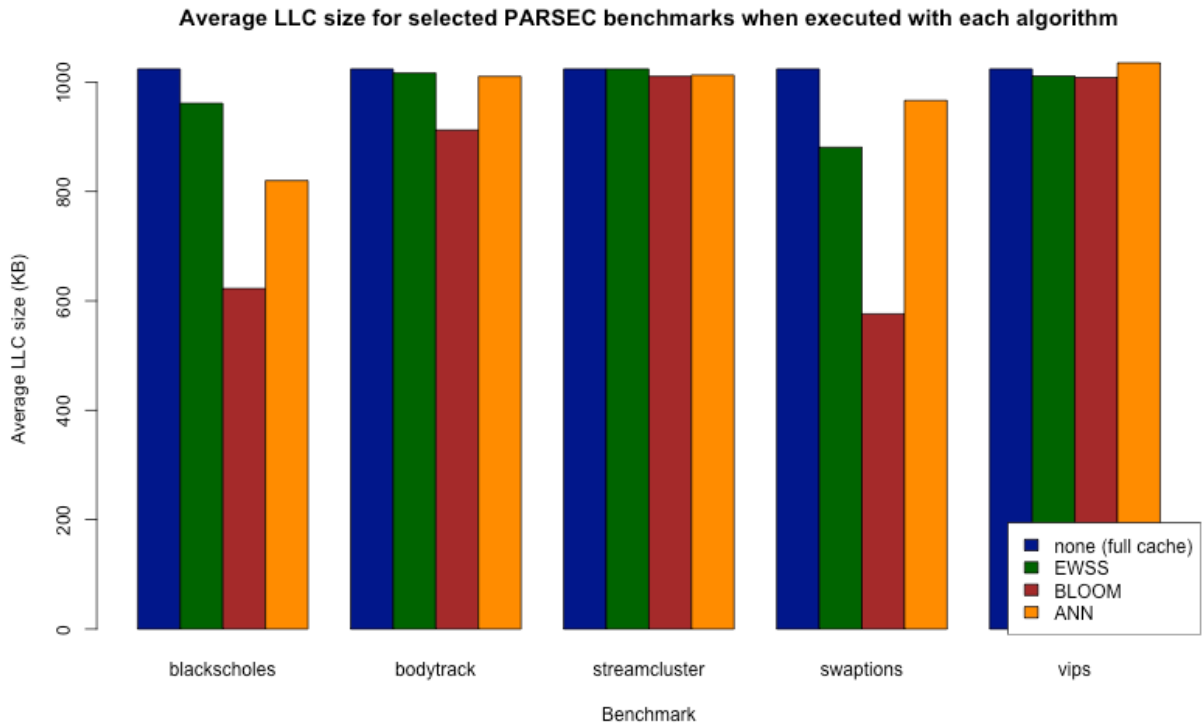

**Figure 12:** The last level cache size averaged over each CPU cycle observed while executing selected benchmarks from the PARSEC suite while using each LLC size selection algorithm presented in this paper.

Figure 12 shows the improvements provided by our method through reduced LLC size during application execution. The BLOOM algorithm turned off an average of 19% of the LLC while the EWSS algorithm only turned off 4.5% of the LLC. The ANN algorithm performed only slightly better than EWSS by turning off an average of 5.4% of the LLC.

Though they both use the same training corpus, the ANN algorithm does not select as small an average cache size as the BLOOM algorithm. One reason for this is that the ANN model is not trained on as many MAS as the BLOOM model is. As shown in Figure 7, duplicate MAS observed across multiple profiles in the training corpus are removed from the training data before FANN creates a neural network. This step is to prevent very common MAS from being mapped to two different optimal LLC sizes, which would prevent the model from representing a function from MAS to optimal LLC size and hence would make the relationship unlearnable by the neural network. This step is unnecessary in the BLOOM algorithm since each bloom filter is disjoint; if the MAS is found in multiple filters the smallest LLC size can always be prefered. Figure 13 shows another informative result relating to the performance repercussions of the amount of data used when training the ANN profiling model.

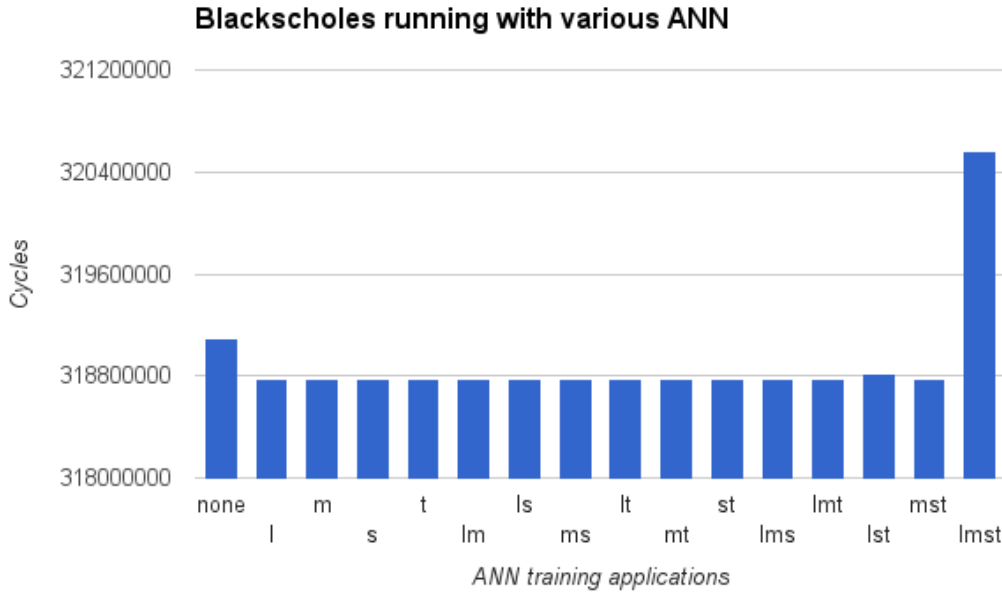

**Figure 13:** Runtime of PARSEC blackscholes benchmark when executed on 4K data with the ANN algorithm using a neural network trained on profiling data from each subset of our application corpus.

The labels on the horizontal axis of Figure 16 are the same as those in the legend of Figure 11. This figure clearly shows a much larger variance in the execution time of the ANN algorithm when trained on profiling data from all 4 applications. This could indicate that the neural network began to overfit the model when exposed to too many examples. Conversely, it could also indicate that too many data points lead to a flatness problem that hindered the ability of the neural network to learn a good partition of the function range. We did not have time to investigate the cause of this variance, so it is left as a question for future research.

## VIII.  HARDWARE COST

To implement the proof-of-concept technique for reducing LLC size presented in this paper some additional hardware costs will be incurred. The LLC requires 2 extra bits per cache block to keep track of read and write accesses so it can be efficiently reconfigured. The LLC will also need 8 bits to maintain the MAS which will need to be updated on each access to the LLC.

The BLOOM algorithm will require one bloom filter for each supported cache size and the number of bits required in each the filter will grow with the size of the training corpus (to accommodate more MAS at the same error rate). To have a 1% chance of collisions will require a set of bloom filters with 4 bits per MAS. In our experiments we added an average of 16,114 MAS into five bloom filters. This would necessitate 40KB of bloom filter memory. The BLOOM algorithm disabled an average of 200KB in our LLC resulting in a total efficiency 160KB of unpowered cache memory. A bloom filter in this configuration will can operate at 5 kHz [24], but our size selection algorithms never block the execution pipeline.

The ANN algorithm requires a neural network with 64 input bits (equal to the length of a MAS) and one output bit for each LLC size. An integrated circuit implementing the neural network will also be required by the ANN algorithm. This neural network is proportional to the number of hardware configurations. In our experiment, we used a fully connected 3 layer neural network with 64 input neurons, 32 hidden neurons, and 5 output neurons. This configuration requires 10 KB of additional memory. The ANN algorithm disabled an average of 46KB in our LLC resulting in a total efficiency 36KB of unpowered cache memory.

## IX. IN GENERAL

We are presenting a general method for using profiling data from a limited corpus of applications to improve performance across all unseen software. Our proof-of-concept implementation of this technique exemplifies the steps needed to apply this method to any domain.

- Design a suitable corpus of applications
- Determine a fixed set of states into which the system will be configurable
- Define a cost function to minimize over the profiling data
- Design a suitable execution signature
- Implement a data model and algorithm to map those signatures to the optimal configuration

In the case of our example, we wrote a corpus of 4 small applications and chose 5 sizes of LLC as the system configurations. The cost function was defined to select the smallest LLC size that did not increase cache misses. We designed memory access signatures to characterize the software/hardware interaction pattern and used MAS to implement a BLOOM and ANN algorithm which predicted optimal cache size using a set of bloom filters or an artificial neural network respectively.

This technique is broadly useful to any environments where novel software is regularly executed. This includes Operating Systems where application software can be downloaded and executed only a single time as well as data centers cannot always guarantee the hardware platform on which an application will execute. Environments using hypervisors or containers to provide PaaS and IaaS would like to completely hide the underlying hardware platform from the application software. This method allows the the provider to mitigate the performance cost of running an application on a hardware platform for which it was not designed without incurring the overhead of profiling the application directly.

## X. CONCLUSIONS AND FUTURE RESEARCH

Traditionally, profiling data must be collected for each new application before it can be used to improve performance. The compulsory unoptimized execution for each first-run application on a new hardware platform creates an unavoidable profiling overhead which scales linearly with the applications to be executed. We presented a method that allows this overhead to be avoided by collecting profiling data only once from a small corpus of applications. We were able to reuse that profiling information across many applications executing for the first time, and we presented a proof-of-concept for adjusting the last-level-cache size which used this profiling data to reduce LLC size by an average of 19% which was 15 percentage points more than the baseline EWSS algorithm.

In this paper, we only considered the case of unknown application software executing on a limited set of hardware configurations defined a priori on which we have exhaustively profiled the applications from our training corpus. Future research may investigate if this technique of using profiling data from a small corpus of applications and hardware configurations can be applied to improving the performance of applications from the training corpus on hardware configurations that were not seen during the training period. We could also evaluate the applicability of this approach to improving the performance of all unknown applications across all unseen hardware configurations. Expanding our method to these use-cases will likely require a larger corpus and research is needed to evaluate the best way to expand the training corpus to generate training data for a wider range of interactions.